\documentclass[runningheads]{llncs}
\usepackage[T1]{fontenc}
\usepackage{graphicx,verbatim}
\usepackage{pifont}
\usepackage{booktabs}
\usepackage{multirow}
\usepackage[]{acronym}
\usepackage{amsmath,amssymb,amsfonts}
\usepackage{wrapfig}
\usepackage{xcolor}
\usepackage{capt-of}
\usepackage{hyperref}
\usepackage{marvosym}

\acrodef{DL}{Deep Learning}
\acrodef{US}{ultrasound}
\acrodef{CNN}{Convolutional Neural Network}
\acrodef{CNNs}{Convolutional Neural Networks}
\acrodef{ROI}{Region of Interest}
\acrodef{SSL}{Semi-Supervised Learning}
\acrodef{GT}{Ground Truth}
\acrodef{CR}{Consistency Regularization}
\acrodef{EMA}{Exponential Moving Average}
\acrodef{MAC}{Mutual Agreement Consistency}
\acrodef{KL}{Kullback–Leibler divergence}
\acrodef{MIG}{Mutual Information Gap}
\acrodef{CE}{Cross-Entropy}
\acrodef{Dice}{Dice based coefficient}
\acrodef{DSC}{Dice Similarity Coefficient}
\acrodef{ASD}{Average Surface Distance}
\acrodef{HD95}{95\% Hausdorff Distance}
\acrodef{SOTA}{state-of-the-art}
\acrodef{PS}{Pubic Symphysis}
\acrodef{FH}{Fetal Head}
\acrodef{NCA}{Neural Cellular Automata}
\acrodef{ViT}{Vision Transformer}
\acrodef{POCUS}{Point-of-Care Ultrasounds}
\acrodef{POC}{point-of-care}
\acrodef{CCA}{Common Carotid Artery}
\acrodef{ReLU}{Rectified Linear Unit}
\acrodef{GELU}{Gaussian Error Linear Unit}
\acrodef{FiLM}{Feature-wise Linear Modulation}
\acrodef{AI}{Artificial Intelligence}
\acrodef{HD}{Hausdorff Distance}
\acrodef{AUC}{Area Under the Curve}
\acrodef{MCC}{Matthews Correlation Coefficient}
\acrodef{MRE}{Mean Radial Error}
\acrodef{IoU}{Intersection over Union}
\acrodef{TS}{task-specific}
\acrodef{AU}{all-task unified}
\acrodef{CG}{clinically-grouped}
\acrodef{ViT}{Vision Transformer}
\acrodef{M2DINO}{Multi-organ and Multi-task DINO framework}
\acrodef{MoE}{Mixture-of-Experts}
\acrodef{DPT}{dense prediction transformer}
\acrodef{OB}{obstetrics}

\begin{document}
\title{Understanding Task Aggregation for Generalizable Ultrasound Foundation Models}
%

\author{
Fangyijie Wang\inst{1,3}$^{\dagger}$\thanks{Corresponding authors: fangyijie.wang@ucdconnect.ie, kathleen.curran@ucd.ie} \and 
Tanya Akumu\inst{2}$^{\dagger}$ \and 
Vien Ngoc Dang\inst{2} \and 
Amelia Jim\'enez-S\'anchez\inst{2} \and 
Jieyun Bai\inst{6,7} \and 
Gu\'enol\'e Silvestre\inst{1,4} 
Karim Lekadir\inst{2,5} \and
Kathleen M. Curran\inst{1,3}$^{\star}$ 
}
\authorrunning{F. Wang et al.}
\institute{
Research Ireland Centre for Research Training in Machine Learning \and
Departament de Matem\`atiques i Inform\`atica, Universitat de Barcelona, Barcelona, Spain \and
School of Medicine, University College Dublin, Dublin, Ireland \and
School of Computer Science, University College Dublin, Dublin, Ireland \and
Instituci\'o Catalana de Recerca i Estudis Avan\c{c}ats (ICREA) \and
Department of Cardiovascular Surgery, The First Affiliated Hospital of Jinan University, Jinan University, Guangzhou, China \and
Auckland Bioengineering Institute, University of Auckland, Auckland, New Zealand \\
$^{\dagger}$ Equal contribution
} 
\maketitle              
\begin{abstract}
Foundation models promise to unify multiple clinical tasks within a single framework, but recent ultrasound studies report that unified models can underperform task-specific baselines. 
We hypothesize that this degradation arises not from model capacity limitations, but from task aggregation strategies that ignore interactions between task heterogeneity and available training data scale. In this work, we systematically analyze when heterogeneous ultrasound tasks can be jointly learned without performance loss, establishing practical criteria for task aggregation in unified clinical imaging models.
We introduce M2DINO, a multi-organ, multi-task framework built on DINOv3 with task-conditioned Mixture-of-Experts blocks for adaptive capacity allocation. We systematically evaluate 27 ultrasound tasks spanning segmentation, classification, detection, and regression under three paradigms: task-specific, clinically-grouped, and all-task unified training.
Our results show that aggregation effectiveness depends strongly on training data scale. While clinically-grouped training can improve performance in data-rich settings, it may induce substantial negative transfer in low-data settings. In contrast, all-task unified training exhibits more consistent performance across clinical groups. We further observe that task sensitivity varies by task type in our experiments: segmentation shows the largest performance drops compared with regression and classification. These findings provide practical guidance for ultrasound foundation models, emphasizing that aggregation strategies should jointly consider training data availability and task characteristics rather than relying on clinical taxonomy alone.

\keywords{Foundation models \and Ultrasound imaging \and Multi-Task learning.}

\end{abstract}
%
%
%


\section{Introduction}

Ultrasound imaging is a cornerstone of clinical care, including obstetrics \cite{Maraci:2020}, cardiology \cite{Villemain:2020}, oncology \cite{Madsen:2011}, and \ac{POC} settings \cite{Self:2022}. It enables rapid, non-invasive, and cost-effective assessment of diverse anatomical structures at the bedside. 
However, ultrasound image appearance varies substantially across operators, devices, and acquisition protocols, complicating robust generalization \cite{Sarris:2012,Huang:2024,Vega:2025}.
Despite recent advances in \ac{DL} for ultrasound, most models focus on isolated task instances (e.g., single-organ segmentation \cite{Kim:2025}, multi-organ classification \cite{Kang:2025}, or multi-organ segmentation \cite{Chen:2025}) or limited combinations of tasks (e.g., joint classification and segmentation \cite{Jiao:2024}), rather than enabling unified multi-organ, multi-task analysis.
Such specialization limits clinical applicability, as real-world workflows require simultaneous multi-organ and multi-task assessment. Foundation models therefore aim to streamline deployment, promote cross-task knowledge sharing, and enable comprehensive ultrasound analysis \cite{Awais:2025}. However, developing a single unified model that reliably performs segmentation, detection, classification, and regression across heterogeneous ultrasound tasks remains an open challenge.

Recent multi-task and foundation-style approaches aim to unify clinical tasks within a single model, thereby simplifying deployment and enabling cross-task knowledge sharing \cite{Jiao:2024,Chen:2025,Kang:2025,Ma:2025}. While these methods report promising results on selected task combinations, a systematic study of how task aggregation strategies influence performance across organs and task types is still lacking. In particular, it remains unclear which tasks can be effectively unified without inducing negative transfer, and how training data scale modulates such interactions.

To address these questions, we introduce \ac{M2DINO}, a DINOv3-based encoder augmented with task-conditio-\\ned \ac{MoE} blocks for large-scale multi-task learning across 27 ultrasound tasks spanning segmentation, classification, detection, and regression. We investigate three training paradigms: (1) \ac{TS} training, where each task is optimized independently; (2) \ac{CG} training, where clinically related tasks are trained jointly; and (3) \ac{AU} training, where all tasks are learned simultaneously within a single model. 

Our contributions are: 
(1) We introduce \ac{M2DINO}, a unified multi-organ, multi-task ultrasound framework built on DINOv3 with task-conditioned \ac{MoE} for adaptive capacity allocation across heterogeneous task objectives.
(2) We introduce a structured framework for evaluating clinical task aggregation and compatibility across organ systems and prediction types.
(3) Through experiments on 27 tasks, we show scale-dependent aggregation effects, identify conditions under which \ac{CG} training induces negative transfer, and provide practical design guidelines for developing unified ultrasound foundation models.

\section{Methodology}

\begin{figure}[t]
\centering
\includegraphics[width=\textwidth]{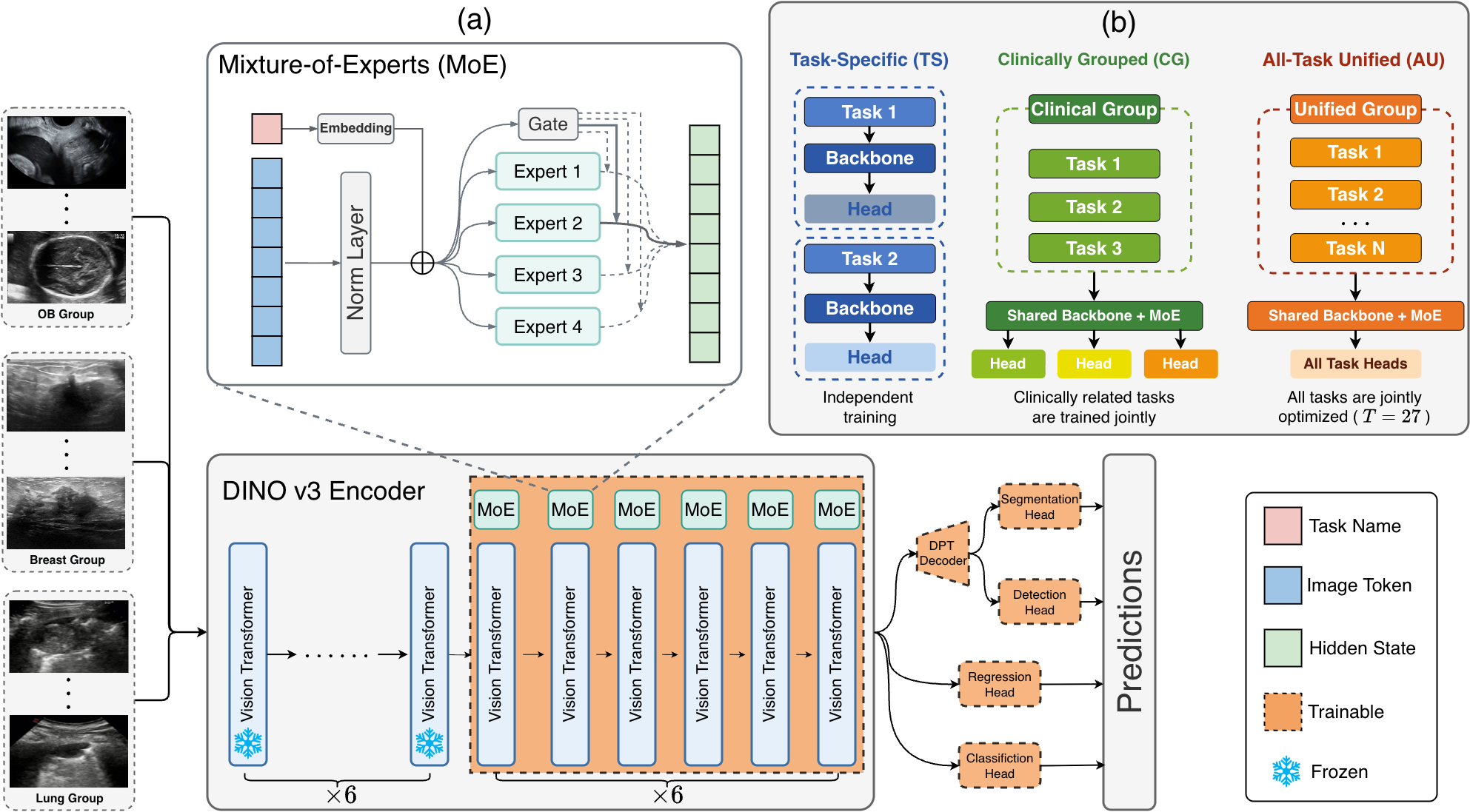}
\caption{Overview of our \ac{M2DINO} framework. {\bf (a)} Ultrasound images are processed by a shared DINOv3 encoder augmented with task-conditioned \ac{MoE} blocks. The unified representation is optimized for segmentation, detection, regression, and classification via task-specific prediction heads. Frozen and trainable components are indicated. {\bf (b)} A conceptual comparison of the three training paradigms. Although the architecture remains the same, \acf{TS}, \acf{CG}, and \acf{AU} differ in how tasks are aggregated during training and in whether the \ac{MoE} is enabled.}
\label{M2DINO}
\end{figure}

This section first formalizes the problem setting and training paradigms (Section~\ref{ssec:training}). We then detail the proposed \ac{M2DINO} architecture, including the backbone~\cite{simeoni:2025}, the task-conditioned \ac{MoE}, and the heads with the multi-task objective (Sections~\ref{ssec:dino}–\ref{ssec:multitask}).
Fig.~\ref{M2DINO} illustrates an overview of the \ac{M2DINO} framework.

\subsection{Problem Setting and Training Paradigms}
\label{ssec:training}

Let $\mathcal{D}=\left\{\mathcal{D}_t\right\}_{t=1}^T$ denote a collection of $T$ ultrasound tasks spanning segmentation, classification, regression, and detection, covering diverse anatomical regions. Each task $\mathcal{D}_t=\left\{\left(\mathbf{X}_i^t, \mathbf{Y}_i^t\right)\right\}$ consists of ultrasound images $\mathbf{X}_i^t$ and \ac{TS} labels $\mathbf{Y}_i^t$. Under the unified training paradigms, our objective is to learn a shared encoder $f_\theta$ that maps an input image $\mathbf{X}$ to a latent representation, which is subsequently optimized by heterogeneous \ac{TS} prediction heads.

We study how different task aggregation strategies affect model performance and transfer behavior within a common DINOv3-based foundation model. Specifically, we evaluate three training paradigms in a controlled comparison setting: 
\begin{itemize}
    \item \textbf{\acf{TS}}: A separate DINOv3 model is trained independently for each task $t$, without any parameter sharing or cross-task interaction.
    \item \textbf{\acf{CG}}: Tasks are jointly trained within predefined clinical groups based on shared organ systems and examination context (e.g., obstetric tasks (OB), breast imaging tasks (Breast), and lung ultrasound tasks (Lung)). Each group shares a DINOv3 \ac{ViT} encoder and task-conditioned \ac{MoE} routing while optimizing heterogeneous prediction objectives (segmentation, classification, detection, or regression).    
    \item \textbf{\acf{AU}}: All $T$ tasks are trained simultaneously within a single shared DINOv3 \ac{ViT} encoder.
\end{itemize}
For unified settings (\ac{CG} and \ac{AU}), the multi-task loss function is defined as: 
$
\mathcal{L}=\sum_{t=1}^T \lambda_t \mathcal{L}_t
$,
where $T$ denotes the number of tasks trained jointly in the current paradigm (e.g., $[3-27]$), and $\mathcal{L}_t$ denotes the \ac{TS} loss and $\lambda_t$ denotes balancing coefficients. Unless otherwise specified, losses are equally weighted. 

In our controlled comparison, all training paradigms share the same pre-trained DINOv3 backbone, \ac{MoE} configuration (when enabled), input resolution, data pre-processing, and optimization settings. As the effective training data size varies across paradigms (e.g., \ac{TS} vs. \ac{AU}), we perform a limited learning rate search within a fixed range for each setting to ensure stable optimization, using a consistent validation-based selection protocol. Table~\ref{tab:experiment_settings} summarizes the experimental settings for each training paradigm.

\begin{table}[t]
\centering
\caption{Experimental settings for evaluating different training paradigms. 
Seg: segmentation; Cls: classification; Reg: regression; Det: detection. MO: Multi-organ.}
\label{tab:experiment_settings}
\scalebox{0.95}{
\setlength{\tabcolsep}{3.5pt}
\begin{tabular}{lcccccccll}
\toprule
\textbf{Training} & \multirow{2}{*}{\textbf{MoE}} & \textbf{\#} & \textbf{\#} & \multirow{2}{*}{\textbf{Reg}} & \multirow{2}{*}{\textbf{Cls}} & \multirow{2}{*}{\textbf{Seg}} & \multirow{2}{*}{\textbf{Det}} & \textbf{Target} & \textbf{Clinical} \\
\textbf{Paradigm} & & \textbf{Tasks} & \textbf{Images} & & & & & \textbf{Anatomy} & \textbf{Group} \\
\midrule
TS & \ding{55} & 1 & 1,144 &  & \ding{51} & & & Breast & Breast \\
TS & \ding{55} & 1 & 1,776 &  & & \ding{51} & & Breast & Breast \\
TS & \ding{55} & 1 & 208 & \ding{51} & & & & Cervical & OB \\
TS & \ding{55} & 1 & 2,818 & \ding{51} & & & & PS/Fetal head & OB \\
TS & \ding{55} & 1 & 762 &  & & \ding{51} & & Fetal abdomen & OB \\
TS & \ding{55} & 1 & 624 & \ding{51} & & & & Fetal femur & OB \\
TS & \ding{55} & 1 & 1,849 &  & & \ding{51} & & Fetal head & OB \\
TS & \ding{55} & 1 & 5,952 &  & \ding{51} & & & Fetal organs & OB \\
TS & \ding{55} & 1 & 483 &  & \ding{51} & & & Fetal breech & OB \\
TS & \ding{55} & 1 & 1,482 &  & & \ding{51} & & Lung & Lung \\
TS & \ding{55} & 1 & 772 &  & \ding{51} & & & Lung & Lung \\
\midrule
CG & \ding{51} & 7 & 11,910 & \ding{51} & \ding{51} & \ding{51} & & Fetal anatomy & OB \\
CG & \ding{51} & 3 & 2,254 & & \ding{51} & \ding{51} & & Lung & Lung \\
CG & \ding{51} & 3 & 2,920 & & \ding{51} & \ding{51} & & Breast & Breast \\
\midrule
AU & \ding{51} & 27 & 32,311 & \ding{51} & \ding{51} & \ding{51} & \ding{51} & MO & All \\
\bottomrule
\end{tabular}
}
\end{table}

\subsection{DINO Backbone}
\label{ssec:dino}

We use the pre-trained DINOv3~\cite{simeoni:2025} model as our encoder backbone. DINOv3 provides \ac{ViT}-S/B/L variants; we adopt the \ac{ViT}-B/16 backbone to balance model capacity with dataset scale and computational efficiency.
Given an ultrasound image, we convert it in RGB format to obtain the input $\mathbf{X} \in \mathbb{R}^{3 \times H \times W}$. The DINOv3-based \ac{ViT} encoder $f_{\theta}$ produces token embeddings $\mathbf{Z}$ and corresponding spatial feature maps $\mathbf{F}$: $(\mathbf{Z}, \mathbf{F})=f_\theta(\mathbf{X})$. Unlike prior multi-task formulations~\cite{Song:2025}, we use spatial feature maps as the unified interface across all tasks. Downstream task-specific heads, including a \ac{DPT} decoder \cite{Ranftl:2021} for segmentation, take the feature maps $\mathbf{F}$ as input. 

Although the $f_{\theta}$ produces token embeddings, we use only the spatial feature maps $\mathbf{F}$ for downstream heads. This design provides a consistent dense feature representation across segmentation, detection, classification, and regression tasks.  Using global token pooling (e.g., the classification token) could favor global prediction tasks over dense prediction tasks such as segmentation and detection. By adopting feature maps $\mathbf{F}$ as the unified interface, we maintain architectural consistency and isolate the effect of task grouping in our compatibility analysis.

\subsection{Mixture of Experts with Task-Conditioned Routing}
\label{ssec:moe}
To mitigate task interference in unified training paradigms (\ac{CG}/\ac{AU}), we integrate task-conditioned \ac{MoE} blocks into the DINOv3 encoder, inspired by \cite{Jain:2023,Lu:2025}. Each task is assigned a unique identifier $t$, which is mapped to a learnable embedding vector: $\mathbf{e}_t=\operatorname{Embedding}(t)$. The gating network (shown in Fig.~\ref{M2DINO}) conditions expert selection on both token embeddings $\mathbf{h}$ and the task embedding $\mathbf{e}_t$:
$
g\left(\mathbf{h}, \mathbf{e}_t\right)=\operatorname{Softmax}\left(W_g\left[\mathbf{h} ; \mathbf{e}_t\right]\right).
$
The output of the \ac{MoE} block is computed as a weighted combination of expert outputs:
$
\mathbf{h}^{\prime}=\sum_{i=1}^K g_i\left(\mathbf{h}, \mathbf{e}_t\right) E_i(\mathbf{h})
$,
where $E_i$ denotes the $i$-th expert and $K$ is the total number of experts. This design enables task-adaptive capacity while maintaining a shared backbone.

Instead of integrating \ac{MoE} into all \ac{ViT} layers, we integrate the \ac{MoE} blocks into the later layers (layers $7-12$, i.e., the last six layers). Early transformer layers tend to encode generic low-level image representations, while later layers encode task-specific representations \cite{Dorszewski:2026}. Restricting \ac{MoE} blocks to later layers enables efficient conditional capacity allocation. Given the scale of our dataset (32,311 training and 8,077 validation samples), we adopt a partial-\ac{MoE} design to balance task-adaptive capacity with computational efficiency.

\subsection{Task-Specific Heads and Multi-Task Learning}
\label{ssec:multitask}

For four different task types, including segmentation, classification, regression, and detection, we develop four lightweight heads to improve computational efficiency. 
Let $\mathbf{F}$ denote the shared feature maps produced by $f_{\theta}$. Each task employs a lightweight prediction head $h_t$ to output $\hat{y}_t = h_t(\mathbf{F})$, where the head parameters are task-specific.
For segmentation tasks, we adopt a \ac{DPT}-style \cite{Ranftl:2021} decoder to generate dense pixel-wise predictions. Classification and regression tasks utilize global pooling followed by fully connected layers, while detection tasks adopt a task-specific detection head. 

Each task is optimized using an appropriate loss function $\mathcal{L}_t$. Specifically, we use Dice loss for segmentation, cross-entropy loss for classification, and $L1$ loss for regression. For detection, we use a single-stage detection loss combining focal loss for pixel-wise supervision and Smooth $L1$ loss for normalized bounding box regression at the corresponding ground-truth center cell. For unified settings (CG/AU), the overall objective is defined in Sec \ref{ssec:training}.

\section{Experiments}

\paragraph{Dataset.}
The dataset is designed to evaluate the model's ability to generalize across four fundamental task categories:
\begin{itemize}
\item \textbf{Segmentation} (12 tasks): Pixel-level annotations for fetal organs (e.g., the head, heart, and abdomen), maternal structures, and lesions. The training set contains 16,615 samples and the test set includes 2,674 samples.
\item \textbf{Classification} (9 tasks): 
Includes fetal standard-plane and fetal position classification, lung disease recognition, and tumor malignancy assessment. The training set has 16,361 samples, and test set has 2,727 samples.
\item \textbf{Detection} (3 tasks): Localization of thyroid nodules, uterine fibroids, and spinal cord injuries (4,333 training / 725 test samples).
\item \textbf{Regression} (3 tasks): Biometric measurements including angle of progression, cervical length, and fetal femur length. The training set includes 3,078 samples, and the test set contains 617 samples.
\end{itemize}
During training, 20\% of the training data are selected for validation.

\paragraph{Implementation Details.}
All methods were trained for 200 epochs with a batch size of 16 using AdamW (initial learning rate $1e-5$, weight decay $1e-4$). The backbone learning rate was set to $2e-5$, the \ac{DPT} head to $1e-5$, \ac{MoE} to $2e-4$, and task-specific heads to $1e-3$ to accelerate convergence. Implementation was based on PyTorch (2.1.2) and Segmentation Models PyTorch \cite{Iakubovskii:2019} with CUDA (12.2), and experiments were conducted on a NVIDIA 4090 GPU. Models were evaluated on the validation set after each epoch, and the best-performing model's weights were saved. Data augmentation and preprocessing followed standard protocols. Full implementation details and code are available at: \href{https://anonymous.4open.science/r/Multitask_Foundation_model_for_ultrasound-2B70}{GitHub}. 

\paragraph{Evaluation Metrics.}
We define standardized evaluation metrics for each of the task types: \textbf{Segmentation}: We report the \ac{DSC} \cite{Dice:1945} for region overlap and \ac{HD}~\cite{Huttenlocher:1993} for boundary accuracy. \textbf{Classification}: We use the \ac{AUC} \cite{Peterson:1954}, F1-score \cite{Sasaki:2015}, and \ac{MCC} \cite{Matthews:1975}. \textbf{Detection}: We use the \ac{IoU}~\cite{zheng:2020} to measure the localization accuracy of predicted bounding boxes. \textbf{Regression}: The \ac{MRE}, reported in pixels, reflects the real-world clinical measurement precision, as it is computed at the original image resolution (i.e., predictions are mapped back from resized inputs).

\section{Results}

\begin{figure}[t]
\centering
\includegraphics[width=\textwidth]{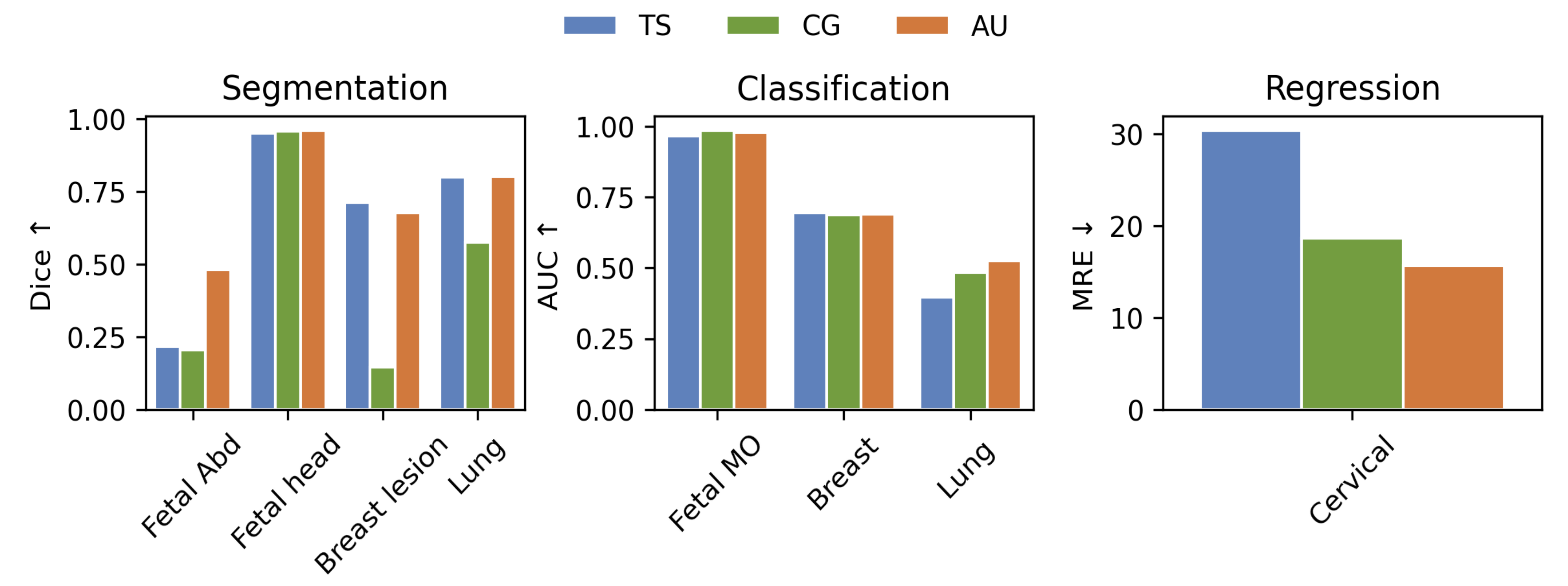}
\caption{Absolute performance of \ac{TS}, \ac{CG}, and \ac{AU} training paradigms across representative tasks: segmentation (\ac{DSC} $\uparrow$), classification (\ac{AUC} $\uparrow$), and regression (\ac{MRE} $\downarrow$). Abd: Abdomen; MO: Multi-organ.
}
\label{fig:barchart}
\end{figure}

Fig. \ref{fig:barchart} presents absolute performance comparisons across training paradigms. In the data-rich \ac{OB} group (11,910 training samples), both \ac{CG} and \ac{AU} training paradigms generally improve over \ac{TS} on most tasks. Specifically, \ac{AU} reduces cervial regression error (\ac{MRE}: 30.4 $\rightarrow$ 15.6) and increases fetal abdomen segmentation overlap (\ac{DSC}: 0.217 $\rightarrow$ 0.481). For fetal head segmentation and multi-organ classification, \ac{CG} and \ac{AU} yield small, modest gains. However, the Breast and Lung groups exhibit different trends. \ac{AU} improves lung classification (\ac{AUC}: 0.396 $\rightarrow$ 0.525). In contrast, \ac{CG} shows large performance drops in breast lesion segmentation (\ac{DSC}: 0.713 $\rightarrow$ 0.145) and lung segmentation (\ac{DSC}: 0.801 $\rightarrow$ 0.576). These results suggest that task aggregation strategies (\ac{CG}/\ac{AU}) benefit from data-rich settings, whereas \ac{CG} is less reliable in low-data settings.

\begin{figure}[tb]
\centering
\begin{minipage}{0.54\linewidth}
    \centering
    \includegraphics[width=.9\linewidth]{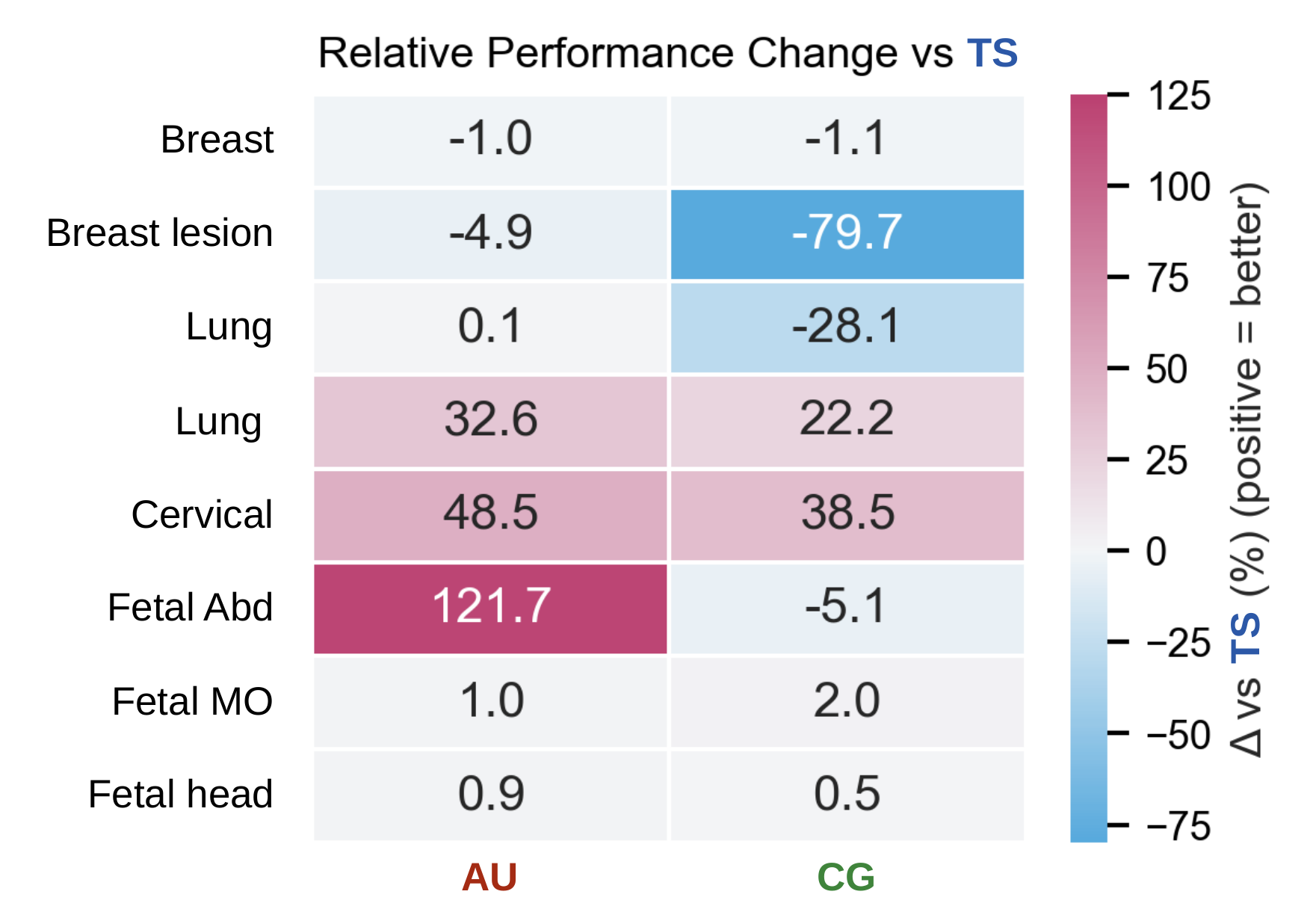}
    \caption{Relative performance change ($\Delta$, \%) with respect to \ac{TS}.}
    \label{fig:heatmap}
\end{minipage}
\hfill
\begin{minipage}{0.45\linewidth}
    \centering
    \captionof{table}{Group-wise average performance change ($\Delta$) relative to \ac{TS}. Positive values indicate improvement; for regression (\ac{MRE}), the sign is adjusted accordingly.
    }
    \label{tab:group_delta}
    \begin{tabular}{lccc}  
    \toprule
    \multirow{2}{*}{\textbf{Group}} & \multirow{2}{*}\textbf{\#} & \multirow{2}{*}{$\boldsymbol{\Delta}$\textbf{(\ac{CG})}} & \multirow{2}{*}{$\boldsymbol{\Delta}$\textbf{(\ac{AU})}} \\
    & \textbf{Images} &  &  \\
    \midrule
    \ac{OB}  & 11,910 & $+\textbf{2.93}$ & $+\textbf{3.76}$ \\
    Breast   & 2,920  & $-0.29$ & $-0.02$ \\
    Lung     & 2,254  & $-0.07$ & $+\textbf{0.07}$ \\
    \bottomrule
    \end{tabular}
\end{minipage}
\end{figure}

To quantify task aggregation effects, Fig. \ref{fig:heatmap} reports relative performance changes with respect to \ac{TS}. It shows that the impact of \ac{CG} and \ac{AU} depends strongly on data scale. In the \ac{OB} group (11,910 training samples), both \ac{AU} and \ac{CG} outperform \ac{TS} across most tasks, with the largest improvements in regression and segmentation. \ac{CG} shows a 5.1\% performance drop in fetal abdomen segmentation. 
In contrast, \ac{CG} shows significant performance drops in smaller groups (Breast and Lung), especially in breast lesion segmentation (-79.7\%). By comparison, \ac{AU} exhibits comparatively less performance changes (-4.9\%).
These results suggest that task aggregation interacts strongly with data availability, and that \ac{CG} is more prone to negative transfer in low-data settings.

Table~\ref{tab:group_delta} summarizes the group-wise average performance change relative to \ac{TS} training. \ac{CG} yields positive improvements in the \ac{OB} ($\Delta=+2.93$; 11,910 samples), but shows slight average decreases in Breast ($\Delta=-0.29$) and Lung ($\Delta=-0.07$). In contrast, \ac{AU} exhibits more stable performance across datasets ($+3.76$ in \ac{OB}, $-0.02$ in Breast, and $+0.07$ in Lung) and generally outperforms \ac{CG} in smaller-scale settings. These results suggest that the effectiveness of \ac{CG} training paradigm depends strongly on data scale.

\section{Discussion}


Our study shows that the effectiveness of task aggregation strategies (\acf{CG}/\acf{AU}) in ultrasound imaging is strongly dependent on training data scale. In the data-rich \acf{OB} group, both \ac{CG} and \ac{AU} improve performance over \acf{TS} (Table~\ref{tab:group_delta}). However, in smaller groups (Breast and Lung), \ac{CG} induces significant negative transfer, indicating that clinical grouping alone does not guarantee positive transfer.

Importantly, \ac{AU} shows more stable performance across groups and fewer large performance drops than \ac{CG} (Fig. \ref{fig:heatmap}). This suggests that broader task aggregation may provide a regularizing effect that reducing overfitting when data are limited. Our findings highlight that partial grouping (i.e., \ac{CG}) can be more prone to negative transfer in small datasets, whereas all-task aggregation yields more reliable transfer behavior.

Furthermore, we observe task-type-dependent effects in our experiments. Segmentation shows the largest performance drops and negative transfer, while regression and classification remain comparatively stable (Fig.~\ref{fig:barchart}). These results suggest that the design of aggregation strategies for foundation models should consider clinical taxonomy together with data scale and task characteristics.

This study has several limitations. 
First, we focus on a single backbone (DINOv3) and predefined clinical grouping strategies. Alternative architectures, such as ultrasound-specific foundation models (e.g., USFM \cite{Jiao:2024} and TinyUSFM \cite{Ma:2025}), or data-driven grouping schemes, may lead to different outcomes.
Second, our analysis is limited to ultrasound imaging. Future work should examine whether similar transfer patterns generalize to other 2D modalities (e.g., radiography or digital pathology) as well as 3D domains such as CT and MRI. Despite these limitations, our findings provide empirical evidence that aggregation strategy and data scale are important factors influencing the performance and stability of unified medical foundation models.

\section{Conclusion}
We present a large-scale empirical analysis of task aggregation strategies for multi-task ultrasound foundation models across 27 heterogeneous clinical tasks. Our findings show that aggregation effectiveness is governed not only by clinical taxonomy but also by data scale and task characteristics. Clinically-grouped aggregation improves performance in data-rich settings but can induce negative transfer in low-data settings. In contrast, anatomy-agnostic aggregation provides more stable cross-task transfer. Segmentation tasks are particularly sensitive to aggregation design, underscoring the need for principled task selection. 
These results demonstrate that naive task scaling does not guarantee improved foundation models and provide practical guidelines for constructing reliable and scalable ultrasound foundation models, with implications for broader medical imaging applications.

\begin{credits}
\subsubsection{\ackname} This work was funded by Taighde \'{E}ireann – Research Ireland through the Research Ireland Centre for Research Training in Machine Learning \\(18/CRT/6183). This research was supported by a grant from the European Research Council (ERC) under the European Union's Horizon Europe research and innovation programme (AIMIX project - Grant Agreement No. 101044779).

\subsubsection{\discintname}
The authors have no competing interests to declare that are relevant to the content of this article. 
\end{credits}

%
%
%
\bibliographystyle{splncs04}
\bibliography{mybibliography}
%




\end{document}